\documentclass[aps, amsmath, amssymb, 10pt,pra,twocolumn, longbibliography, superscriptaddress]{revtex4-1}

\usepackage{algorithm}
\usepackage{algorithmic}
\usepackage{amsmath}
\usepackage{amsthm}
\usepackage{bm}
\usepackage{braket}
\usepackage{subfigure}
\usepackage[dvipdfmx]{graphicx}
\usepackage{float}
\usepackage{url}
\usepackage{xcolor}
\usepackage{multirow}
\usepackage[colorlinks=true,urlcolor=blue,citecolor=blue,linkcolor=blue]{hyperref}
\usepackage{cleveref}
\usepackage{here}

\crefname{figure}{Fig.}{Figs.}
\crefname{equation}{Eq.}{Eqså.}
\crefname{section}{Sec.}{Sec.}
\Crefname{figure}{Figure}{Figures}
\Crefname{equation}{Equation}{Equations}
\Crefname{section}{Section}{Sections}

\begin{document}

%
%

\title{Cross cross resonance gate}

\author{Kentaro Heya}
\affiliation{IBM  Quantum, IBM Research Tokyo,  19-21  Nihonbashi  Hakozaki-cho,  Chuo-ku,  Tokyo,  103-8510,  Japan}
\affiliation{Research Center for Advanced Science and Technology (RCAST), The University of Tokyo, Meguro-ku, Tokyo 153-8904, Japan}
\email{kheya@qc.rcast.u-tokyo.ac.jp}

\author{Naoki Kanazawa}
\affiliation{IBM  Quantum, IBM Research Tokyo,  19-21  Nihonbashi  Hakozaki-cho,  Chuo-ku,  Tokyo,  103-8510,  Japan}
\email{knzwnao@jp.ibm.com}

%
%

\begin{abstract}

Implementation of high-fidelity swapping operations is of vital importance to execute quantum algorithms on a quantum processor with limited connectivity.
We present an efficient pulse control technique, cross-cross resonance (CCR) gate, to implement iSWAP and SWAP operations with dispersively-coupled fixed-frequency transmon qubits.
The key ingredient of the CCR gate is simultaneously driving both of the coupled qubits at the frequency of another qubit, wherein the fast two-qubit interaction roughly equivalent to the XY entangling gates is realized without strongly driving the qubits.
We develop the calibration technique for the CCR gate and evaluate the performance of iSWAP and SWAP gates
The CCR gate shows roughly two-fold improvement in the average gate error and more than 10~\% reduction in gate times from the conventional decomposition based on the cross resonance gate.

\end{abstract}

\maketitle

%
%

\section{Introduction}

The number of qubits available in a superconducting quantum processor has been continuously growing despite the restricted connectivity exemplified by square or heavy-hexagon lattice structure~\cite{hertzberg2020laserannealing}.
These connections are carefully designed to avoid overlapping of transition frequencies with neighboring qubits, while enabling the implementation of one of the error correction codes which is the key ingredient of the fault-tolerant quantum processors~\cite{takita2017experimental,kelly2015state,corcoles2015demonstration,barends2014superconducting,takita2016demonstration,arute2019quantum}.
Though such processors can implement state-of-the-art two-qubit entangling gates with average gate error reaching below $10^{-2}$ \cite{jurcevic2021demonstration}, insertion of multiple swapping operations to entangle non-connected qubit pairs becomes a visible obstacle in building large scale quantum processors.

In superconducting qubits, entangling gates are generally implemented using electric dipole interactions between qubits, which are roughly divided into two subclasses.
One of the implementations is realized by dynamically modulating device parameters to switch the coupling with the aid of tunable elements~\cite{blais2003tunable, hime2006solid, niskanen2007quantum, stehlik2021tunable}.
The additional control lines for the tunable elements may induce an extra degree of interaction with a noisy environment and such architecture tends to show shorter coherence times.
Another class of implementation can be illustrated as manipulation of Hamiltonian terms by irradiating a pair of dispersively coupled qubits with microwave pulses \cite{rigetti2010fully, PhysRevLett.109.240505, Chow_2013, PhysRevApplied.14.044039, Noguchi_2020}.
Such architectures can be realized without tunable elements and thus qubits can be well protected from the source of decoherence at the cost of longer gate time impeded by the crosstalk error due to the static ZZ interaction~\cite{Ku_2020, kandala2020demonstration}.

The cross resonance (CR) gate is a typical implementation of the latter subclass without any tunable element, and this simple architecture benefits the device scale-up \cite{rigetti2010fully, Chow_2011}.
On the other hand, CR gate tends to have a longer gate time, i.e. about a few hundred nanoseconds depending on the power of microwave drive, especially when the echo sequence is incorporated \cite{Sheldon_2016}.
In principle, the stronger the drive power, the shorter the gate time.
However, the strong microwave drive breaks the underlying approximations, leading to leakage to higher energy levels outside of the computational subspace \cite{Malekakhlagh_2020, Tripathi_2019}.
The CR gate can implement gates in the XX gate family represented by the CNOT gate, and any two-qubit gate operation can be expressed with at most 3 CNOT gates.
The SWAP gate consists of 3 CNOT gates and this is an essential operation to execute quantum circuits requiring arbitrary qubit connection.
In spite of its importance, the three-fold longer gate time can drastically deteriorate the performance of quantum circuits and system-level metrics such as quantum volume~\cite{Cross_2019}.

In this paper, we present a pulse sequence, the cross-cross resonance~(CCR) gate, which enables more efficient iSWAP and SWAP gates implementation without spoiling the advantage of the longer coherence time of dispersively-coupled fixed-frequency transmon qubits.
This control scheme implements iSWAP and SWAP gates with 2 and 3 entangling gates, respectively.
While the total number of required two-qubit gates is the same as the conventional decomposition based on the CR gate, the CCR gate has a higher interaction speed and shorter gate time.
The calibration of CCR gate is scrutinized, and we introduce the channel purification technique enabling the investigation of an approximated unitary representation of experimental gates.
The calibrated CCR gate demonstrates remarkable improvements in the average gate error which are confirmed by the interleaved randomized benchmarking~\cite{EAZ2005, KLRetc2008, MGE2011}.
These experiments were done using Qiskit Pulse through the IBM Quantum cloud provider \cite{Alexander_2020, garion2020experimental}.

%
%

\section{Principle}
We describe the model of the CCR gate using the Hamiltonian of standard cQED setup.
A system consisting of two qubits dispersively coupled to a bus resonator can be modeled with Duffing oscillators
\begin{align}
\mathcal{H}=
& \sum_{i\in[0,1]} \left\{\tilde{\omega}_i b_{i}^\dagger b_{i} + \frac{\alpha_i}{2}b_{i}^\dagger b_{i}^\dagger b_{i} b_{i}\right\} \\ \nonumber
& + g\left(b_{0}^\dagger + b_{0}\right) \left(b_{1}^\dagger + b_{1}\right) \\ \nonumber
& + \sum_{i\in[0,1]} \Omega_i\cos({\omega_{di}} t)(b_{i}^\dagger + b_{i}),
\end{align}
where $b_{i}$ ($b_{i}^\dagger$) is annihilation (creation) operator with $\tilde{\omega_{i}}$, $\alpha_i$ and $g$ being the dressed eigenfrequency, anharmonicity of $i$-th qubit, and the coupling between two qubits, respectively.
The last term describes the general drive Hamiltonian for each qubit with the time-invariant drive $\Omega_i$ and the drive frequency $\omega_{di}$.
Let $i \in [0, 1]$ denote the label for the control and target qubit, respectively.

\subsection{Cross resonance~(CR) gate} \label{sec:cr_model}

We start with the Hamiltonian of the CR gate, which is the predecessor of our proposal.
The CR gate is a microwave-only two-qubit entangling gate that is commonly used for fixed-frequency dispersively-coupled qubits \cite{Chow_2011}.
In the CR gate, the control qubit is irradiated with a microwave tone at the frequency of the target qubit $\omega_{d0} = \tilde{\omega}_1 \sim \omega_1 - g^2/\Delta$ with $\Delta = \tilde{\omega}_0 - \tilde{\omega}_1$ being the detuning of two qubits.
This stimulus induces a controlled rotation on the target qubit whose direction depends on the state of the control qubit.

For simplicity, we model the transmons as ideal qubits
Ignoring the classical crosstalk~\cite{Sheldon_2016}, the effective Hamiltonian of the CR gate for a qubit model under the rotating wave approximation $\Omega_0/(\tilde{\omega}_0 + \tilde{\omega}_1) \ll 1$, the strong dispersive condition $g/\Delta \ll 1$, and the weak drive condition $\Omega/\Delta \ll 1$ is written as follows~\cite{Magesan_2020}
\begin{align}
\mathcal{H}_\mathrm{eff}
\sim
\frac{\mbox{sgn}(\Delta)g\Omega_0}{\sqrt{\Delta^2+\Omega_0^2}}\frac{ZX}{2},
\label{eq:cr_eff_ham}
\end{align}
where ZX denotes the tensor product of the standard single-qubit Pauli operator Z and X.
Note that the above notation of the effective Hamiltonian $\mathcal{H}_{\mathrm{eff}}$ is a representation in a rotating frame moved from the laboratory frame using the unitary operator
\begin{align}
R=e^{-i\frac{\Omega_0^2}{2\Delta}t\frac{ZI}{2}}.
\end{align}

To clarify the difference between the CR and CCR gates, we characterize the two-qubit gate implemented by each pulse sequence with the KAK decomposition~\cite{e15061963}.
Any two-qubit gate is described as $4\times4$ matrix in $\mathcal{SU}(4)$, which can be decomposed as follows,
\begin{align}
U=k_1 A k_2,
\end{align}
where $k_1$ and $k_2$ are $4\times4$ matrices in $\mathcal{SU}(2)\otimes\mathcal{SU}(2)$ and $A$ is a $4\times4$ matrix in the maximal Abelian subgroup $\mathcal{A}$ of $\mathcal{SU}(4)$.
The matrix $A$ is parameterized as follows,
\begin{align}
A(c_1, c_2, c_3)=\exp{\left(-i\sum_{i=1}^{3} c_i \frac{\sigma_i \otimes \sigma_i}{2}\right)},
\end{align}
where $c_{1,2,3}\in[0,\pi/2]$ are the Cartan coefficients and $\sigma_i \in \left\{ X, Y, Z \right\}$ is one of the single-qubit Pauli operators.
Typical two-qubit gates such as CNOT, iSWAP, and SWAP gate have $(\pi/2,0,0)$, $(\pi/2,\pi/2,0)$, and $(\pi/2,\pi/2,\pi/2)$ as Cartan coefficients, respectively.
These coefficients explain why the SWAP gate requires the overhead of $3$ CNOT gates.
The CR gate also has the Cartan coefficients as follows,
\begin{align}
\bm{c}(t)
=
\left(
\frac{\mbox{sgn}(\Delta)g\Omega_0}{\sqrt{\Delta^2+\Omega_0^2}}t,
0,
0
\right),
\end{align}
which is locally equivalent to the controlled-rotation gate, namely fundamentals of the CNOT gate.
Note that without qubit approximation, terms derived from the higher-order levels are added to the effective Hamiltonian~\cite{Magesan_2020}.

\subsection{Cross cross resonance~(CCR) gate} \label{sec:ccr_model}
The CCR gate is a natural extension of the CR gate, where the control and target qubit are simultaneously irradiated with microwave tones at the frequency of each other; $\omega_{d0}=\tilde{\omega_1}-\Omega_0^2/2\Delta$ and $\omega_{d1}=\tilde{\omega_0}+\Omega_1^2/2\Delta$.
In contrast to the CR gate, drive frequencies of the CCR gate are the function of drive amplitudes, $\Omega_0$ and $\Omega_1$, owing to the drive-induced Stark shift~\cite{gambetta2006qubit, Schneider_2018}.
We again use the qubit model for simplicity, and we find the effective Hamiltonian of the CCR gate can be described as follows,
\begin{align}
\mathcal{H}_{\mathrm{eff}}\sim
\frac{\mbox{sgn}(\Delta)g\Omega_0}{\sqrt{\Delta^2 + \Omega_0^2 + 2\Omega_1^2}}
\frac{ZX}{2}
-
\frac{\mbox{sgn}(\Delta)g\Omega_1}{\sqrt{\Delta^2 + 2\Omega_0^2 + \Omega_1^2}}
\frac{XZ}{2}.
\end{align}
Given $\Omega_1=0$, the effective Hamiltonian of the CCR gate matches that of the CR gate shown in Eq \eqref{eq:cr_eff_ham}.
Details are shown in~\cref{sec:effective_hamiltonian}.
The CCR gate has the Cartan coefficients as follows,
\begin{align}
\bm{c}(t)
=
\left(
\frac{+\mbox{sgn}(\Delta)g\Omega_0}{\sqrt{\Delta^2+\Omega_0^2+2\Omega_1^2}}t,
\frac{-\mbox{sgn}(\Delta)g\Omega_1}{\sqrt{\Delta^2+2\Omega_0^2+\Omega_1^2}}t,
0
\right).
\end{align}
As mentioned in \cref{sec:ccr_rb}, we can generate the SWAP gate with $3$ CCR gates.

Though the SWAP gate can be implemented based on $3$ CR gates or $3$ CCR gates, the gate time required for entanglement generation with limited pulse amplitudes depends on the implementation.
\begin{align}
t_{\mathrm{SWAP}}^{\mathrm{CR}}
&=\frac{3\pi\sqrt{\Delta^2+\Omega^2}}{2g\Omega}, \
t_{\mathrm{SWAP}}^{\mathrm{CCR}}
&=\frac{3\pi\sqrt{\Delta^2+\frac{3}{4}\Omega^2}}{2g\Omega},
\label{eq:ccr_catran}
\end{align}
where both CCR drive amplitudes are set to half of the CR drive amplitude $\Omega$ for the sake of fair comparison between two entangling gates.
Because $\Delta^2, \Omega^2 > 0$, we find $t_{\mathrm{SWAP}}^{\mathrm{CR}} > t_{\mathrm{SWAP}}^{\mathrm{CCR}}$, hence, the CCR gate can implement faster swapping operations.

Note that these discussions are based on the effective Hamiltonian for a qubit model.
Relaxing this approximation yields quantum crosstalk and leakage from the computational basis, but we expect the CCR gate can alleviate these errors as a consequence of weak drives.

%
%

\begin{figure*}[!t]
	\begin{center}
		\includegraphics[width=\textwidth]{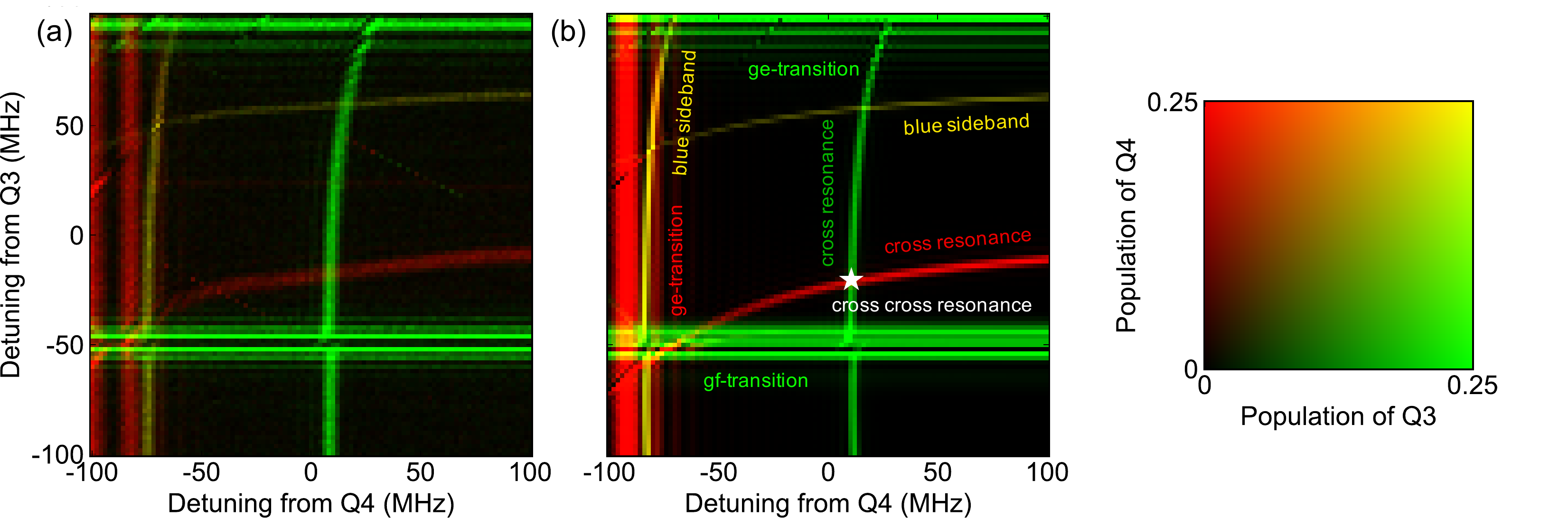}
		\caption{
		Heatmaps of qubit excited state population measured after CCR gates with various frequency detunings from the predetermined drive frequencies based on the individual CR gate calibration.
		Experimental result (a) and simulation result (b).
		The measured population of Q3 and Q4 are overlaid in the same plot to highlight the optimal driving point.
		The excited state population of Q3 and Q4 correspond to the red and green, respectively.
		The yellow-colored region indicates the frequencies where both Q3 and Q4 are simultaneously excited.
		See discussions in the main text.
		In each panel, the vertical (horizontal) axis represents the detuning from the predetermined CR frequency of Q4 (Q3) corresponding to the frequency of target qubit Q3 (Q4).
		The labels in the figure~(b) indicate the origin of each transition.
		}
		\label{double_freq_sweeping}
	\end{center}
\end{figure*}

\section{Experiment}
\subsection{System}
The experiments presented in this work are executed on the Q3 and Q4 out of the five-qubit IBM Quantum system \texttt{ibmq\_bogota}.
This processor consists of dispersively-coupled fixed-frequency transmon qubits, and the control electronics are capable of generating arbitrary waveforms with a cycle time $\mathrm{dt} = 0.222~\mathrm{ns}$ afforded by a fast digital to analog converter.
The device parameters at the time of the experiment are shown in~\cref{device_parameter}.
\begin{table}[h]
	\caption{Parameters of qubits}
	\label{device_parameter}
	\begin{tabular}{p{2cm}p{2cm}p{2cm}} \hline \hline
	                    &   Q3                      &   Q4                  \\ \hline
		$\omega/2\pi$   & 	4.858~GHz               &   4.978~GHz               \\
		$\alpha/2\pi$ 	&  	-324~MHz                &   -338~MHz                \\
		$T_1$           &  112.4~$\mathrm{\mu s}$   &   115.5~$\mathrm{\mu s}$   \\
		$T_2$           &  191.7~$\mathrm{\mu s}$   &   167.6~$\mathrm{\mu s}$  \\ \hline \hline
	\end{tabular}
\end{table}

The coupling strength between two qubits was found to be $g/2\pi=1.40~\mathrm{MHz}$.
Because the CCR gate drives cross resonance from both directions, the drive Hamiltonian consists of multiple frequency components.
To realize a high fidelity CCR gate, it is important that all transition frequencies, including higher energy levels, are well isolated \cite{hertzberg2020laserannealing}.
The selected qubit pair satisfies this requirement, thus it is considered to be a suitable experimental platform for our demonstration.

First, we independently calibrate a CR pulse with Q3 as control and Q4 as target qubit, and then vice versa.
The CR pulses correspond to the $R_{\mathrm{ZX}}(\pi/4)$ and $R_{\mathrm{XZ}}(\pi/4)$ gates, respectively.
Both of the CR pulse envelopes are \emph{GaussianSquare} pulses, i.e. a square pulse with Gaussian-shaped rising and falling edges.
And both of the pulses have total duration $t_{\mathrm{CR}}=856~\mathrm{dt}=190.2~\mathrm{ns}$.
The square portion of the pulses have a duration of $600~\mathrm{dt}$ and the Gaussian rising and falling edges last $128~\mathrm{dt}$ and have a $64~\mathrm{dt}$ standard deviation.
In \texttt{ibmq\_bogota} a standard CR gate is calibrated with the rotary echo, which eliminates almost all unwanted dynamics by driving the target qubit with an intense microwave tone at the resonance frequency.
However, this technique is not applicable to the CCR gate because induced local Hamiltonian terms $\mathrm{IX+XI}$ may suppress generator terms $\mathrm{XZ+ZX}$ of the gate.
In this regard, we calibrate CR gates with the active cancellation, wherein a target qubit is driven by a weak resonance tone in the opposite phase to the unwanted local Hamiltonian dynamics $\mathrm{IX+IY}$~\cite{Sheldon_2016}.
The CR gates are carefully tuned with the selective Pauli error amplifying technique which we recently developed~\cite{heya2021pauli}, and this is akin to the Hamiltonian error amplifying tomography~(HEAT) technique~\cite{Sundaresan_2020}.
We should keep in mind that unwanted dynamics represented by a non-local ZZ interaction may still alive and slightly contribute to the CCR gate dynamics.
See \cref{sec:anti_cross} for detailed analysis.

\subsection{Calibration of drive frequencies}

Next, we calibrate the detuning of each microwave tone frequency to realize the CCR gate by fixing the tone amplitude at the optimal point of underlying CR gates.
As mentioned in \cref{sec:ccr_model}, the existence of higher-order levels changes the amount of the frequency shifts due to the off-resonance microwave drives.
Therefore, we need to simultaneously calibrate two drive frequencies $\omega_{d0}$ and $\omega_{d1}$.
In what follows, we denote qubit $i = 0$ (1) to 3 (4) to confirm the physical qubit index.
Note that we apply an active cancellation tone for another CR drive along with a CR tone for the qubit, which usually differs by frequency $\sim\Delta/(2\pi)$.

\Cref{double_freq_sweeping} shows the experimental~(a) and numerical~(b) excited state population of qubits after the simultaneous excitation while sweeping the drive frequencies, where the excited state population refers to the probability of existence other than the ground state.
In the numerical simulation, we treat a transmon as a three-level system and apply parameters shown in \cref{device_parameter}, which is clarified by the preliminary experiments.
The experimental result is in qualitatively good agreement with the theoretical prediction.
The optimal detuning frequencies are highlighted by a star symbol where two CR transitions crossover.
This point is also experimentally confirmed at $(\tilde{\omega}_3' - \tilde{\omega}_3)/2\pi = -18.0~\mathrm{MHz}$, $(\tilde{\omega}_4' - \tilde{\omega}_4)/2\pi = 8.9~\mathrm{MHz}$.

In addition to the CR dynamics, several unwanted transitions are also confirmed by the experiment.
The g-f transition of Q4 indicates a frequency on resonance with the non-linear two-photon transition process $\left|gg\right\rangle \rightarrow \left|gf\right\rangle$ occurring at $\tilde{\omega}_4 + \alpha_4 / 2$ which is roughly $-50~\mathrm{MHz}$ apart from $\tilde{\omega}_3$.
The g-e transition $\left|gg\right\rangle \rightarrow \left|ge\right\rangle$ of this qubit is found at $\tilde{\omega}_4 \sim \tilde{\omega}_3 + \Delta$.
The g-e transition $\left|gg\right\rangle \rightarrow \left|eg\right\rangle$ of Q3 is also observed at $\tilde{\omega}_3 \sim \tilde{\omega}_4 - \Delta$.
We also find the two-photon blue sideband transition, in which $\left|gg\right\rangle \rightarrow \left|ee\right\rangle$ is driven at $\tilde{\omega}_3 + \tilde{\omega}_4 / 2$, and this appears as yellow trajectories~\cite{Leek_2009}.
The optimal driving point is sufficiently isolated from these transitions, and this indicates we can generate a stable CCR gate with Q3 and Q4.
See \cref{sec:double_freq} for the details.

\subsection{Calibration of gate time} \label{sec:cartan_sweep}
As previously discussed, it is not efficient to control the time evolution of the CCR gate with pulse amplitude because calibrated optimal tone frequencies get offset due to the AC Stark effect.
Thus, we calibrate the target gate by scanning over the CCR gate time to find the optimal value, at which we realize an entangling gate \emph{almost} locally equivalent to the $\sqrt{\mathrm{iSWAP}}$ gate which is a perfect entangler~\cite{e15061963}.
Owing to the nonzero ZZ interaction, a single CCR gate does not generate an entangling gate completely locally equivalent to iSWAP-type gates.
However, as will be described later in~\cref{sec:ccr_rb}, this interaction can be nullified with multiple CCR gates composing the iSWAP or SWAP gate.
See \cref{sec:anti_cross} for the details.

\begin{figure}
	\begin{center}
		\includegraphics[width=0.4\textwidth]{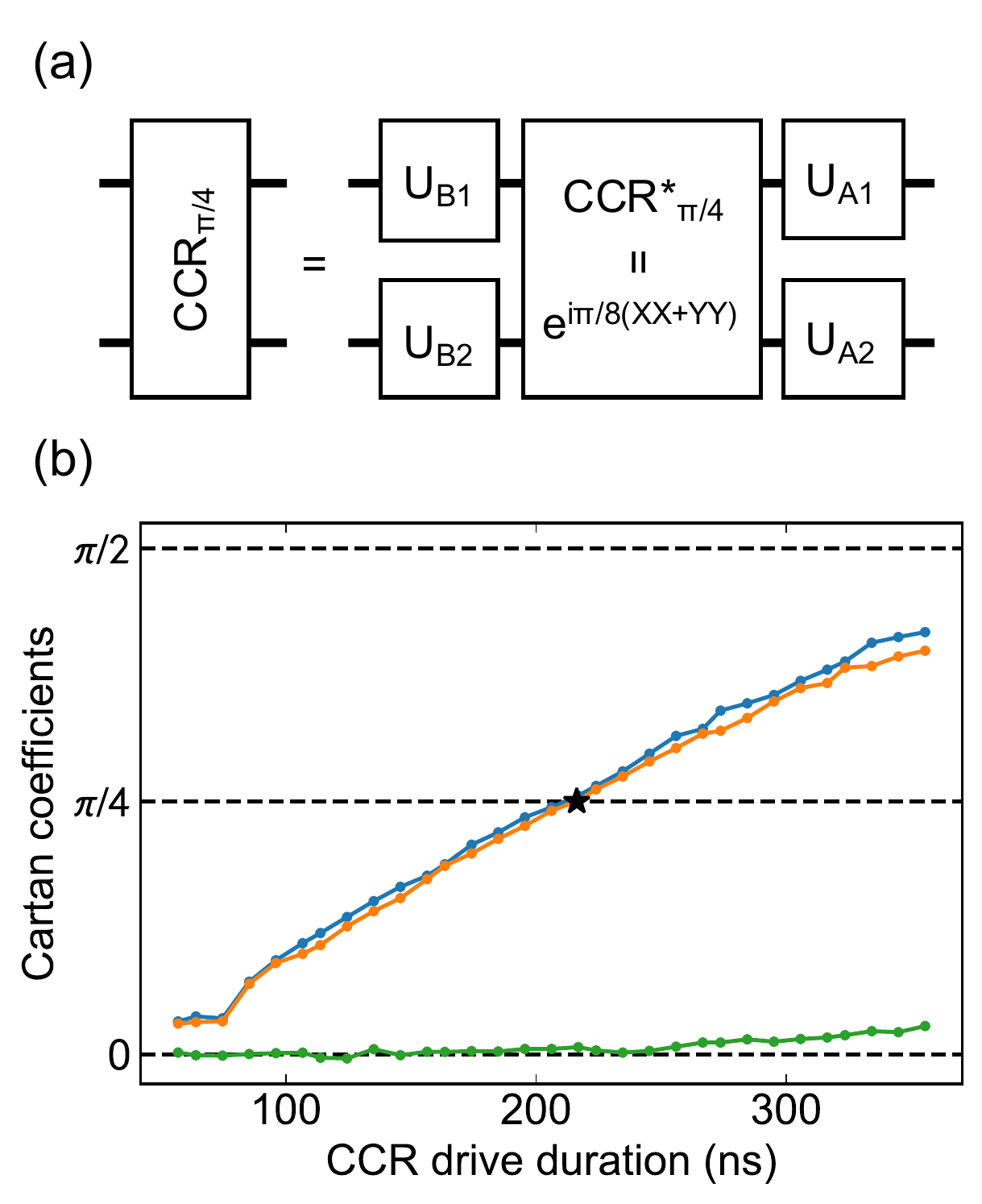}
		\caption{
		(a) Desired condition for the entangling gate generated by the CCR drive.
		(b) Approximated Cartan coefficients of the entangling gates while sweeping the CCR drive duration.}
		\label{cartan_time_sweep}
	\end{center}
\end{figure}

The target CCR gate is found when the following conditions for Cartan coefficients are satisfied,
\begin{align}
c_1(t) + c_2(t) &= \frac{\pi}{2} \\ \nonumber
c_3(t) = 0.
\label{eq:cartan}
\end{align}
To this end, we need to measure the Cartan coefficients while sweeping the CCR gate time.
The evolution of a quantum system can be investigated with the quantum process tomography~(QPT) which extracts a superoperator representation of quantum process with tomographic reconstruction~\cite{Mohseni_2008}.
Because the Cartan coefficients are only defined on the unitary matrix representation, we extract a unitary matrix from a superoperator estimated by the QPT experiment with the aid of the channel purification protocol.
This technique is detailed in~\cref{sec:channel_purification}.
\Cref{cartan_time_sweep}~(b) shows the approximated Cartan coefficients of the entangling gates while sweeping the CCR drive total duration from $256~\mathrm{dt}$ to $1606~\mathrm{dt}$ with the fixed rise and fall edges.
It can be found that 2 Cartan coefficients increase at the same rate and one remains at zero as predicted by our effective Hamiltonian model shown in \cref{eq:ccr_catran}, and this indicates that the CCR drive generates the entangling gates in the XY interaction family wherein the iSWAP belongs~\cite{Abrams_2020}.
From \Cref{cartan_time_sweep}~(b), we find the optimal CCR drive total duration at $981~\mathrm{dt}=218.0~\mathrm{ns}$.

As shown in \Cref{cartan_time_sweep}~(a), the unitary matrix corresponding to the CCR drive with optimal time can be decomposed as follows,
\begin{align}
\mathrm{CCR}_{\pi/4}=
\left(U_{\mathrm{A1}}\otimes U_{\mathrm{A2}}\right)
\mathrm{CCR}^*_{\pi/4}
\left(U_{\mathrm{B1}}\otimes U_{\mathrm{B2}}\right),
\end{align}
where $U_{\mathrm{A1,A2,B1,B2}}$ represent $2\times2$ matrices and $\mathrm{CCR}^*_{\pi/4}$ represents a $4\times4$ matrix in the maximum Abelian subgroup of $\mathcal{SU}(4)$.

\subsection{Two-qubit randomized benchmarking with the CCR gates} \label{sec:ccr_rb}

\begin{figure}
	\begin{center}
		\includegraphics[width=0.4\textwidth]{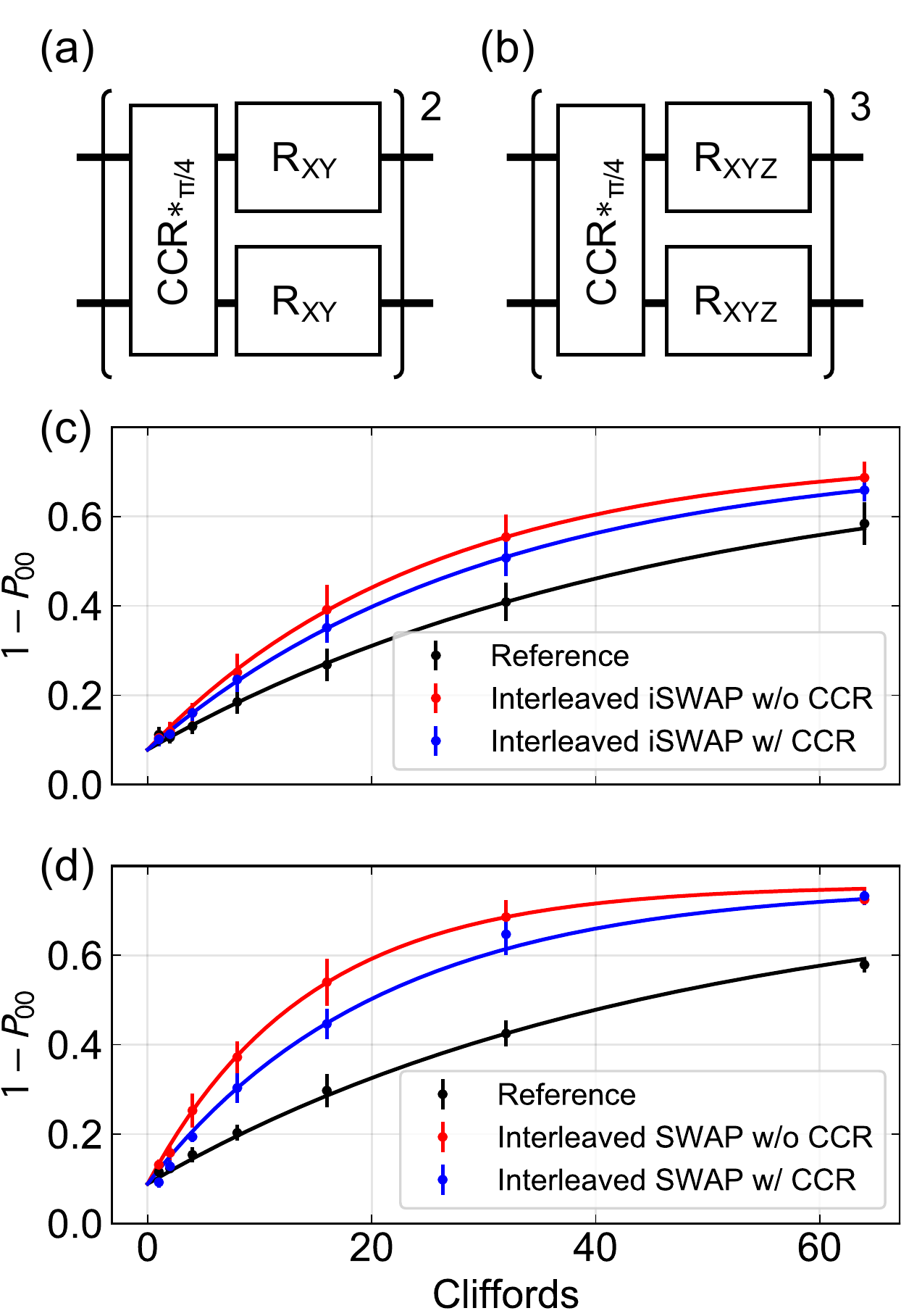}
		\caption{
		(a) Gate sequence to generate the iSWAP gate with twice CCR gates.
		(b) Gate sequence to generate the SWAP gate with three times CCR gates.
		Experimental results of the two-qubit interleaved randomized benchmarking for the iSWAP~(c) and SWAP~(d) gate with and without the CCR gates.
		}
		\label{two_qubit_rb}
	\end{center}
\end{figure}

Finally, we implement the iSWAP and SWAP gates consisting of multiple CCR gates and compare the performance of these gates with the standard gate decomposition based on the CR gates.
We have already obtained the expression of the local rotations $U_{\mathrm{A1,A2,B1,B2}}$ by the KAK decomposition of $\mathrm{CCR}_{\pi/4}$.
Therefore, We can generate the $\mathrm{CCR}^*_{\pi/4}$ gates as follows,
\begin{align}
\mathrm{CCR}^*_{\pi/4} =
\left(U^\dagger_{\mathrm{A1}}\otimes U^\dagger_{\mathrm{A2}}\right)
\mathrm{CCR}^*_{\pi/4}
\left(U^\dagger_{\mathrm{B1}}\otimes U^\dagger_{\mathrm{B2}}\right).
\end{align}
As shown in the \cref{two_qubit_rb}~(a) and (b), we can implement the iSWAP and SWAP gates with the echo sequences as follows,
\begin{align}
U_{\mathrm{iSWAP}}&=
\left(
\mathrm{CCR}^*_{\pi/4}~
R_{\mathrm{XY}}
\otimes
R_{\mathrm{XY}}
\right)^2, \\
U_{\mathrm{SWAP}}&=
\left(
\mathrm{CCR}^*_{\pi/4}~
R_{\mathrm{XYZ}}
\otimes
R_{\mathrm{XYZ}}
\right)^3,
\end{align}
where $R_{\mathrm{XY}}=e^{i\frac{\pi}{2}\frac{X+Y}{\sqrt{2}}}$ and $R_{\mathrm{XYZ}}=e^{i\frac{\pi}{3}\frac{X+Y+Z}{\sqrt{3}}}$ represent the single-qubit gates.
Conventionally the iSWAP and SWAP gates are decomposed into 2 and 3 two-pulse echoed CX~(TPCX) gates respectively, which is an echoed version of the CR gate~\cite{takita2017experimental}.
In our setup, the TPCX gate is calibrated at a gate time of $334.2~\mathrm{ns}$.
On the other hand, the CCR gate is calibrated at $218.0~\mathrm{ns}$ which is restricted by the slow interaction speed from Q3 to Q4.

\begin{table}[h]
	\caption{Comparison of performances of iSWAP and SWAP gates implemented with CR and CCR gate.}
	\begin{tabular}{p{2.5cm}p{1.3cm}p{1.3cm}p{1.3cm}p{1.3cm}} \hline \hline
	  & \multicolumn{2}{c}{iSWAP} & \multicolumn{2}{c}{SWAP} \\
	  & CR           & CCR           & CR           & CCR          \\ \hline
	Gate time~(ns) & 775.1 & 647.1 & 1073.8 & 935.1 \\
	Error ($10^{-2}$) & $1.4(1)$ & $0.8(2)$ & $3.5(9)$ & $2.0(6)$ \\ \hline \hline
	\end{tabular}
	\label{tab:fidelities}
\end{table}

Because these gates belong to the Clifford group, the performance of the gates can be precisely measured with the two-qubit interleaved randomized benchmarking~(IRB) experiment.
\Cref{two_qubit_rb}~(c) and (d) shows the experimental results of the IRB.
We took 10 random circuits for each Clifford sequence length and had 1024 samplings of measurements for each random circuit to get a data point.
The curves of the population other than the ground state are fit by the exponential decay with the same scale and offset parameters in each figure.
The performance of these gates is also summarized in \cref{tab:fidelities}.
This result demonstrates a 42.8~\% reduction in the average gate error of both gates, along with a 16.5~\% and 12.9~\% reduction in the gate time for the iSWAP and SWAP gates, respectively.

%
%

\section{Summary and Discussion}

We have proposed and demonstrated the cross cross resonance (CCR) gate, which is a two-qubit control scheme for dispersively-coupled fixed-frequency transmon qubits.
In the CCR gate, two qubits are simultaneously driven at the frequency of another qubit, and this operation implements both ZX and XZ term in the effective Hamiltonian.
This effective Hamiltonian realizes the target entangling operation in a moderate microwave power regime, in which the leakage error is not significant in spite of its faster gate speed.
At equivalent ZX and XZ interaction strength, the CCR gate generates the entangling gate locally equivalent to the iSWAP rotation.

We have also experimentally shown the calibration of high fidelity CCR gates and composed iSWAP and SWAP gates based on the CCR gate.
By using the interleaved randomized benchmarking, we obtained the gate error of $0.8(2) \times 10^{-2}$ at the gate time of $647.1~\mathrm{ns}$ for iSWAP gate, and $2.0(6) \times 10^{-2}$ at the gate time of $935.1~\mathrm{ns}$ for SWAP gate.
This indicates a 42.8~\% reduction in gate error and more than 10~\% reduction in gate speed by comparing with the conventional implementation based on cross resonance gates.

This work shows an important achievement for the speed-up of two-qubit entangling gates, which will pave the way for further device scaling with restricted connectivity.

%
%

\section*{Acknowledgements}\label{sec:acknowledgements}
We acknowledge fruitful discussions with Emily Pritchett, Ken X. Wei, Petar Jurcevic, Ikko Hamamura, Akhil Pratap Singh, Yutaka Tabuchi, Shuhei Tamate, Atsushi Noguchi, and Yasunobu Nakamura.
We acknowledge people creating and supporting the \texttt{ibmq\_bogota} system on which all data presented here was taken.

\bibliography{manuscript}

\appendix

%
%

%
%

\section{Effective Hamiltonian model of the cross cross resonance for a qubit model}\label{sec:effective_hamiltonian}
In the qubit model, the anharmonicity is infinite so the qubit subspace is perfectly isolated and Hamiltonian is given by
\begin{align}
\mathcal{H}=
&
\omega_0\frac{ZI}{2}
+ \omega_1\frac{IZ}{2}
+ gXX \nonumber\\
&
+ \Omega_0 \cos\left(\omega_{d0} t\right) XI
+ \Omega_1 \cos\left(\omega_{d1} t\right) IX,
\end{align}
where for simplicity we assume a constant amplitude drive $\Omega_i$ on the X quadrature of the $i$-th qubit.
Moving into the frame rotating at $\omega_p=(\omega_{d0}+\omega_{d1})/2$ via the unitary
\begin{align}
R=e^{-i\omega_p t \left(\frac{ZI}{2} + \frac{IZ}{2}\right)},
\end{align}
gives
\begin{align}
\mathcal{H}=
&
(\omega_0 - \omega_p)\frac{ZI}{2}
+ (\omega_1 - \omega_p)\frac{IZ}{2}
+ g\left(\frac{XX}{2}+\frac{YY}{2}\right) \nonumber\\
&
+ \Omega_0
\left(
\cos\left(\omega_m t\right) \frac{XI}{2} - \sin\left(\omega_m t\right) \frac{YI}{2}
\right) \nonumber\\
&
+ \Omega_1
\left(
\cos\left(\omega_m t\right) \frac{IX}{2} + \sin\left(\omega_m t\right) \frac{IY}{2}
\right),
\end{align}
where $\omega_m=(\omega_{d0} - \omega_{d1})/2$.
The Schrieffer-Wolff transformation~\cite{PhysRev.149.491} via the skew-Hermitian operator
\begin{align}
S=-i\frac{g}{\Delta}\left(\frac{XY}{2}-\frac{YX}{2}\right)
\end{align}
gives the block-diagonal Hamiltonian as follows,
\begin{align}
\mathcal{H}\sim
&
(\tilde{\omega_0} - \omega_p)\frac{ZI}{2}
+ (\tilde{\omega_1} - \omega_p)\frac{IZ}{2} \nonumber\\
&
+ \Omega_0
\left(
\cos\left(\omega_m t\right) \frac{XI}{2} - \sin\left(\omega_m t\right) \frac{YI}{2}
\right) \nonumber\\
&
+ \Omega_1
\left(
\cos\left(\omega_m t\right) \frac{IX}{2} + \sin\left(\omega_m t\right) \frac{IY}{2}
\right) \nonumber\\
&
+ \frac{g\Omega_0}{\Delta}
\left(
\cos\left(\omega_m t\right) \frac{ZX}{2} - \sin\left(\omega_m t\right) \frac{ZY}{2}
\right) \nonumber\\
&
- \frac{g\Omega_1}{\Delta}
\left(
\cos\left(\omega_m t\right) \frac{XZ}{2} + \sin\left(\omega_m t\right) \frac{YZ}{2}
\right),
\end{align}
where $\tilde{\omega_0}=\omega_0 + g^2/\Delta$ and $\tilde{\omega_1}=\omega_1 - g^2/\Delta$.
Note that we truncated terms included in $O((g/\Delta)^3)$.
Next, to take into account the drive induced frequency shift, moving into the rotating frame via the unitary
\begin{align}
R=e^{-i\omega_m t \left(\frac{ZI}{2} - \frac{IZ}{2}\right)}
\end{align}
gives
\begin{align}
\mathcal{H}\sim
&
(\tilde{\omega_0} - \omega_{d0}) \frac{ZI}{2}
+ (\tilde{\omega_1} - \omega_{d1}) \frac{IZ}{2} \nonumber\\
&
+ \Omega_0 \frac{XI}{2}
+ \Omega_1 \frac{IX}{2} \nonumber\\
&
+ \frac{g\Omega_0}{\Delta}
\left(
\cos\left(2\omega_m t\right) \frac{ZX}{2} - \sin\left(2\omega_m t\right) \frac{ZY}{2}
\right) \nonumber\\
&
- \frac{g\Omega_1}{\Delta}
\left(
\cos\left(2\omega_m t\right) \frac{XZ}{2} + \sin\left(2\omega_m t\right) \frac{YZ}{2}
\right).
\end{align}
The Schrieffer-Wolff transformation via the skew-Hermitian operator
\begin{align}
S=
&
-i\left(
\frac{\Omega_0}{\Delta_0} \frac{YI}{2}
+
\frac{\Omega_1}{\Delta_1} \frac{IY}{2}
\right),
\end{align}
where $\Delta_i=\tilde{\omega_i}-\omega_{di}$, gives the block-diagonal Hamiltonian as follows,
\begin{align}
\mathcal{H}\sim
&
\Delta_0 \left(1+\frac{r_0^2}{2}\right)\frac{ZI}{2}+\Delta_1 \left(1+\frac{r_1^2}{2}\right)\frac{IZ}{2} \nonumber\\
&
-\frac{\Omega_0 r_0^2}{3}\frac{XI}{2} -\frac{\Omega_1 r_1^2}{3}\frac{IX}{2}
\nonumber\\
&
+\left\{\frac{g\Omega_0}{\Delta}\left(1-\frac{r_0^2+r_1^2}{2}\right) + \frac{g\Omega_1}{\Delta}r_0 r_1\right\}\cos(2\omega_m t)\frac{ZX}{2}
\nonumber\\
&
-\left\{\frac{g\Omega_1}{\Delta}\left(1-\frac{r_0^2+r_1^2}{2}\right) + \frac{g\Omega_0}{\Delta}r_0 r_1\right\}\cos(2\omega_m t)\frac{XZ}{2}
\nonumber\\
&
+ \frac{g}{\Delta}\left(r_1\Omega_1-r_0\Omega_0\right)\cos(2\omega_m t)\frac{XX}{2}
\nonumber\\
&
+ \frac{g}{\Delta}\left(r_1\Omega_0-r_0\Omega_1\right)\cos(2\omega_m t)\frac{ZZ}{2}
\nonumber\\
&
-\frac{g\Omega_0}{\Delta}\left(1-\frac{r_0^2}{2}\right)\sin(2\omega_m t)\frac{ZY}{2}
\nonumber\\
&
-\frac{g\Omega_1}{\Delta}\left(1-\frac{r_1^2}{2}\right)\sin(2\omega_m t)\frac{YZ}{2}
\nonumber\\
&
+\frac{g\Omega_0}{\Delta}r_0\sin(2\omega_m t)\frac{XY}{2}
\nonumber\\
&
+\frac{g\Omega_1}{\Delta}r_1\sin(2\omega_m t)\frac{YX}{2},
\end{align}
where $r_i=\Omega_i/\Delta_i$.
Note that we truncated terms included in $O((\Omega_0/\Delta)^4)$ and $O((\Omega_1/\Delta)^4)$.
We set the microwave drive frequencies $\omega_{di}$ to
\begin{align}
\omega_{d0}&=\omega_1-\frac{\Omega_0^2}{2\Delta}\\
\omega_{d1}&=\omega_0+\frac{\Omega_1^2}{2\Delta}
\end{align}
so as to satisfy the following conditions
\begin{align}
\Delta_0 \left(1+\frac{r_0^2}{2}\right)
=
-\Delta_1 \left(1+\frac{r_1^2}{2}\right)
=
-2\omega_m,
\end{align}
and we can rewrite the Hamiltonian as follows,
\begin{align}
\mathcal{H}\sim
&
\Delta\left\{1+\frac{1}{2}\left(\frac{\Omega_0}{\Delta}\right)^2\right\} \frac{ZI}{2}
-
\Delta\left\{1+\frac{1}{2}\left(\frac{\Omega_1}{\Delta}\right)^2\right\} \frac{IZ}{2} \nonumber\\
&
-\frac{\Omega_0}{3}\left(\frac{\Omega_0}{\Delta}\right)^2 \frac{XI}{2}
-\frac{\Omega_1}{3}\left(\frac{\Omega_1}{\Delta}\right)^2 \frac{IX}{2} \nonumber\\
&
+\frac{g\Omega_0}{\Delta}\left\{1-\frac{1}{2}\left(\frac{\Omega_0}{\Delta}\right)^2 - \frac{3}{2}\left(\frac{\Omega_1}{\Delta}\right)^2\right\}\cos(2\omega_m t)\frac{ZX}{2}\nonumber\\
&
-\frac{g\Omega_1}{\Delta}\left\{1-\frac{3}{2}\left(\frac{\Omega_0}{\Delta}\right)^2 - \frac{1}{2}\left(\frac{\Omega_1}{\Delta}\right)^2\right\}\cos(2\omega_m t)\frac{XZ}{2}\nonumber\\
&
-g\left\{\left(\frac{\Omega_0}{\Delta}\right)^2 + \left(\frac{\Omega_0}{\Delta}\right)^2\right\}\cos(2\omega_m t)\frac{XX}{2} \nonumber\\
&
-2g\frac{\Omega_0}{\Delta}\frac{\Omega_1}{\Delta}\cos(2\omega_m t)\frac{ZZ}{2} \nonumber\\
&
-\frac{g\Omega_0}{\Delta}\left\{1-\frac{1}{2}\left(\frac{\Omega_0}{\Delta}\right)^2\right\}\sin(2\omega_m t)\frac{ZY}{2}\nonumber\\
&
-\frac{g\Omega_1}{\Delta}\left\{1-\frac{1}{2}\left(\frac{\Omega_1}{\Delta}\right)^2\right\}\sin(2\omega_m t)\frac{YZ}{2}\nonumber\\
&
+g\left(\frac{\Omega_0}{\Delta}\right)^2\sin(2\omega_m t)\frac{XY}{2} \nonumber\\
&
+g\left(\frac{\Omega_1}{\Delta}\right)^2\sin(2\omega_m t)\frac{YX}{2}.
\end{align}
Finally, moving into the rotating frame via the unitary
\begin{align}
R=
e^{i2\omega_m t
\left(
\frac{ZI}{2}-\frac{IZ}{2}
\right)},
\end{align}
gives
\begin{align}
\mathcal{H}\sim
&
-\frac{\Omega_0}{3}\left(\frac{\Omega_0}{\Delta}\right)^2
\left\{\cos(2\omega_m t)\frac{XI}{2}- \sin(2\omega_m t)\frac{YI}{2}\right\} \nonumber\\
&
-\frac{\Omega_1}{3}\left(\frac{\Omega_1}{\Delta}\right)^2
\left\{\cos(2\omega_m t)\frac{IX}{2} + \sin(2\omega_m t)\frac{IY}{2}\right\} \nonumber\\
&
+\frac{g\Omega_0}{\Delta}\left\{1-\frac{1}{2}\left(\frac{\Omega_0}{\Delta}\right)^2-\left(\frac{\Omega_1}{\Delta}\right)^2\right\}\frac{ZX}{2} \nonumber\\
&
-\frac{g\Omega_0}{\Delta}\left(\frac{\Omega_1}{\Delta}\right)^2 \left\{\cos(4\omega_m t)\frac{ZX}{2} + \sin(4\omega_m t)\frac{ZY}{2}\right\} \nonumber\\
&
-\frac{g\Omega_1}{\Delta}\left\{1-\left(\frac{\Omega_0}{\Delta}\right)^2-\frac{1}{2}\left(\frac{\Omega_1}{\Delta}\right)^2\right\}\frac{XZ}{2} \nonumber\\
&
-\frac{g\Omega_1}{\Delta}\left(\frac{\Omega_0}{\Delta}\right)^2 \left\{\cos(4\omega_m t)\frac{XZ}{2} - \sin(4\omega_m t)\frac{YZ}{2}\right\} \nonumber\\
&
-2g\frac{\Omega_0}{\Delta}\frac{\Omega_1}{\Delta}\cos(2\omega_m t)\frac{ZZ}{2} \nonumber\\
&
-g\left\{\left(\frac{\Omega_0}{\Delta}\right)^2+\left(\frac{\Omega_0}{\Delta}\right)^2\cos(4\omega_m t)\right\}\cos(2\omega_m t)\frac{XX}{2} \nonumber\\
&
-g\left(\frac{\Omega_1}{\Delta}\right)^2\cos(4\omega_m t)\sin(2\omega_m t)\frac{XY}{2} \nonumber\\
&
+g\left\{\left(\frac{\Omega_0}{\Delta}\right)^2\sin(2\omega_m t)\right. \nonumber\\
&\left. +\left(\frac{\Omega_1}{\Delta}\right)^2\cos(2\omega_m t)\sin(4\omega_m t)\right\}\frac{YX}{2} \nonumber\\
&
+g\left(\frac{\Omega_1}{\Delta}\right)^2\sin(4\omega_m t)\sin(2\omega_m t)\frac{YY}{2}.
\end{align}
The above Hamiltonian $\mathcal{H}$ can be divided into a stationary term and a perturbation with a period $T=\pi/\omega_m$, and we can calculate the effective Hamiltonian from the Magnus expansion as follows,
\begin{align}
\mathcal{H}_{\mathrm{eff}}\sim
\frac{\mbox{sgn}(\Delta)g\Omega_0}{\sqrt{\Delta^2 + \Omega_0^2 + 2\Omega_1^2}}
\frac{ZX}{2}
-
\frac{\mbox{sgn}(\Delta)g\Omega_1}{\sqrt{\Delta^2 + 2\Omega_0^2 + \Omega_1^2}}
\frac{XZ}{2}.
\end{align}
Note that the above notation of the effective Hamiltonian $\mathcal{H}_{\mathrm{eff}}$ is a representation on a rotating frame moved from the laboratory frame using the unitary operator
\begin{align}
R=
e^{-i\frac{\Omega_0^2}{2\Delta}t\frac{ZI}{2}}
e^{i\frac{\Omega_1^2}{2\Delta}t\frac{IZ}{2}}.
\end{align}

%
%

\section{CCR drive frequencies with the various drive amplitudes}\label{sec:double_freq}

\begin{figure*}
	\begin{center}
		\includegraphics[width=\textwidth]{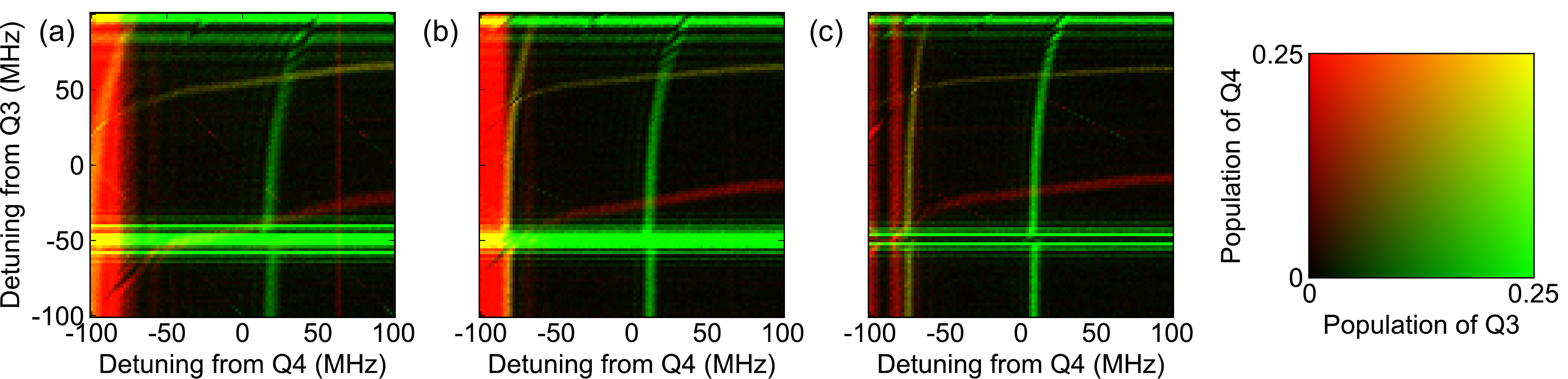}
		\caption{
		Heatmaps of qubit excited state population measured after CCR gates with various frequency detunings from the predetermined drive frequencies based on the individual CR gate calibration.
		The measured population of Q3 and Q4 are overlaid in the same plot.
		In each panel, the vertical (horizontal) axis represents the detuning from the predetermined CR frequency of Q4 (Q3) corresponding to the frequency of target qubit Q3 (Q4).
		The total duration of the $R_{\mathrm{ZX}}(\pi/4)$ gate is set to (a) $145.8$, (b) $168.0$, and (c) $190.2~\mathrm{ns}$.
		}
		\label{double_freq_sweeping_2}
	\end{center}
\end{figure*}

While the CCR drive, the eigenfrequencies of both qubits are shifted by the drive-induced AC Stark effect.
Here we observe the dependence of the CCR drive frequencies operating point on the CCR drive strength.
In this experiment, first, we independently calibrate the $R_{\mathrm{ZX}}(\pi/4) $ and $R_{\mathrm{XZ}}(\pi/4)$ gate by the CR gates with different gate time in $145.8, 168.0$, and $190.2~\mathrm{ns}$.
Note that the pulse amplitude and gate time are inversely proportional.
Next, while maintaining the respective drive amplitudes, the CR drives from both directions are applied at the same time with various drive frequencies.
The \cref{double_freq_sweeping_2} illustrates the eigenfrequency of Q3 and Q4 are the function of the gate time, which are redshifted and blueshifted with respect to the drive amplitude, respectively.

In \cref{double_freq_sweeping_2}~(a), it can be seen that the cross resonance from Q4 to Q3 and the gf-transition of the Q4 are overlapped.
It suggests that the CR drive irradiating to Q3 causes the population leakage from the first excited state of Q3 to the second excited state of Q4.
This phenomenon is not peculiar to the CCR gates but also occurs in the ordinal CR gates.
We find that the two-dimensional drive frequency sweep experiment will bring important information to cope with these leakage errors.

%
%

\section{Channel purification}\label{sec:channel_purification}

To analyze the approximate Cartan coefficients of the experimentally generated quantum gate, we introduce the channel purification method which is derived from the McWeeny purification method~\cite{doi:10.1098/rspa.1956.0100}.
In the channel purification method, representation of quantum channel $\mathcal{E}$ is first translated into a Choi-representation according to the channel-state duality, as follows,
\begin{align}
\rho_{\mathrm{choi}}&=\sum_{ij}\chi_{ij} \ket{P_i}\bra{P_j},
\end{align}
where $\ket{P_i}$ is a vectorized $i$-th Pauli operator and $\chi_{ij}$ is a coefficient of the corresponding Chi-matrix of the channel.
Here, Chi-matrix is calculated as follows,
\begin{align}
\mathcal{E}(\rho) = \sum_{ij}\chi_{ij}P_i \rho P_j.
\end{align}
Next, we project the Choi-density matrix as follows,
\begin{align}
\rho_{\mathrm{choi}}\rightarrow \ket{\psi_{\mathrm{max}}}\bra{\psi_{\mathrm{max}}},
\end{align}
where $\ket{\psi_{\mathrm{max}}}$ is the eigenvector of $\rho_{\mathrm{choi}}$ with the maximum eigenvalue.
The projected Choi-density matrix is translated back to the Kraus-representation according to the channel-state duality.
When the Choi-density matrix is projected to rank-$1$, there is only one Kraus operator~$\mathrm{K}$, and such a quantum channel satisfies complete positive but not trace-preserving.
To derive a unitary operator that approximates the Kraus operator, we project the Kraus operator as follows,
\begin{align}
\mathrm{K}
&=M_{\mathrm{L}}\mathrm{diag}\left\{r_i e^{i\theta_i}\right\}M_{\mathrm{R}} \\
&\rightarrow M_{\mathrm{L}}\mathrm{diag}\left\{\theta_i\right\}M^{-1}_{\mathrm{L}}
\end{align}
where $M_{\mathrm{L},\mathrm{R}}$ are regular matrix and the $r_i e^{i\theta_i}$ is a complex number.

Recently, a similar method to reconstruct a unitary matrix representation from a quantum channel has been introduced in Ref.~\cite{sugiyama2020reliable}.

%
%

\section{Anti-crossing of the Cartan coefficients}\label{sec:anti_cross}

\begin{figure*}
	\begin{center}
		\includegraphics[width=\textwidth]{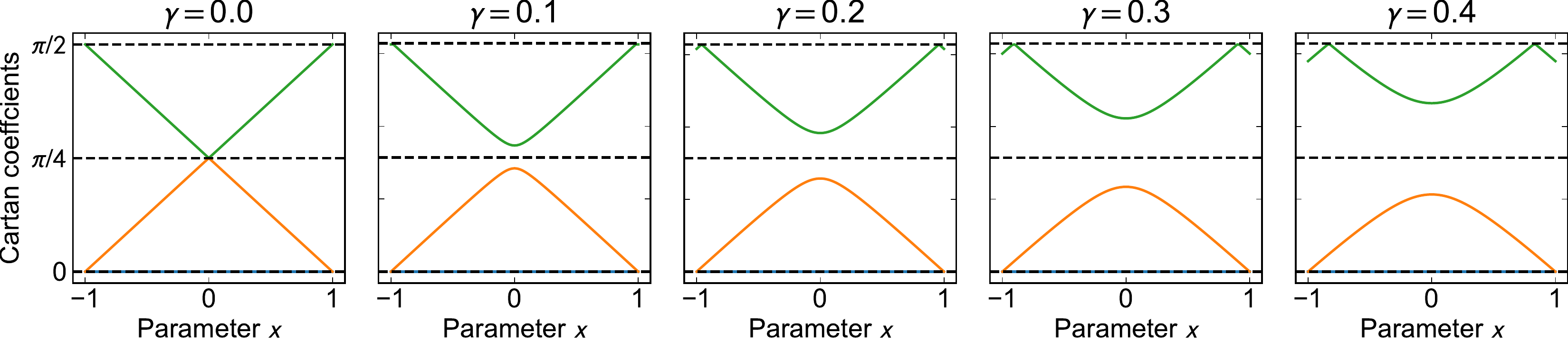}
		\caption{
		Cartan coefficients of the unitary gate generated by the CCR drive with the static ZZ error $\gamma ZZ~(\gamma\in[0,0.4])$, while sweeping the ratio between the bidirectional CR drive amplitudes as $(1-x):(1+x)~(x\in[-1,1])$.
		Each line color corresponds to a Cartan coefficient, namely there are three lines for $c_1$, $c_2$ and $c_3$.
		}
		\label{cartan_anti_cross}
	\end{center}
\end{figure*}

As mentioned in \cref{sec:cartan_sweep}, iSWAP gates cannot be implemented with a single CCR gate though Eq. \eqref{eq:ccr_catran} shows two nonzero Cartan coefficients.
Here we qualitatively estimate the imperfection of the CCR gate with the simplified Hamiltonian model.
As discussed in \cref{sec:ccr_model}, the effective Hamiltonian of the CCR gate consists of the ZX and XZ terms arising from the bidirectional cross resonance drive.
In addition, it is thought that the IX, IY, XI, and YI terms will arise from classical crosstalk, and the ZZ term arises when the higher-order excited levels are taken into account.
In principle, the terms derived from the classical crosstalk can be eliminated by adding active cancellation drives.
Therefore, we model the effective Hamiltonian as follows
\begin{align}
H=
\left(\frac{1-x}{2}\right)\frac{XZ}{2} + \left(\frac{1+x}{2}\right)\frac{ZX}{2} +
\gamma \frac{ZZ}{2},
\end{align}
where $x$ is a parameter corresponding to the ratio between the bidirectional CR drive amplitudes, $\gamma$ is a perturbation coefficient for the static ZZ term.
We numerically calculated the Cartan coefficient for the unitary gate $U=e^{i\pi/2 H}$ with various $x$, $\gamma$.
\Cref{cartan_anti_cross} shows the results of the numerical simulation, where we calculated within $x\in[-1,1]$ and $\gamma\in[0,0.4]$.
Recalling that the iSWAP-type gate can be represented by Cartan coefficients $c_1 = c_2,$ and $c_3 = 0$, the nonzero ZZ interaction, which is a dominant error source without rotary echo~\cite{gambetta2006qubit}, induces an anti-crossing of two Cartan coefficients yielding $c_1 \neq c_2$.
This simulation indicates a high fidelity iSWAP cannot be implemented with a single CCR gate with qubits with finite ZZ error.
On the other hand, suppression of the ZZ interaction~\cite{kandala2020demonstration} enables us to implement the iSWAP gate with a single CCR gate and remarkably shortens the gate time.

%
%

\section{Pulse sequences of iSWAP gate}\label{sec:waveform}

\begin{figure*}
	\begin{center}
		\includegraphics[width=\textwidth]{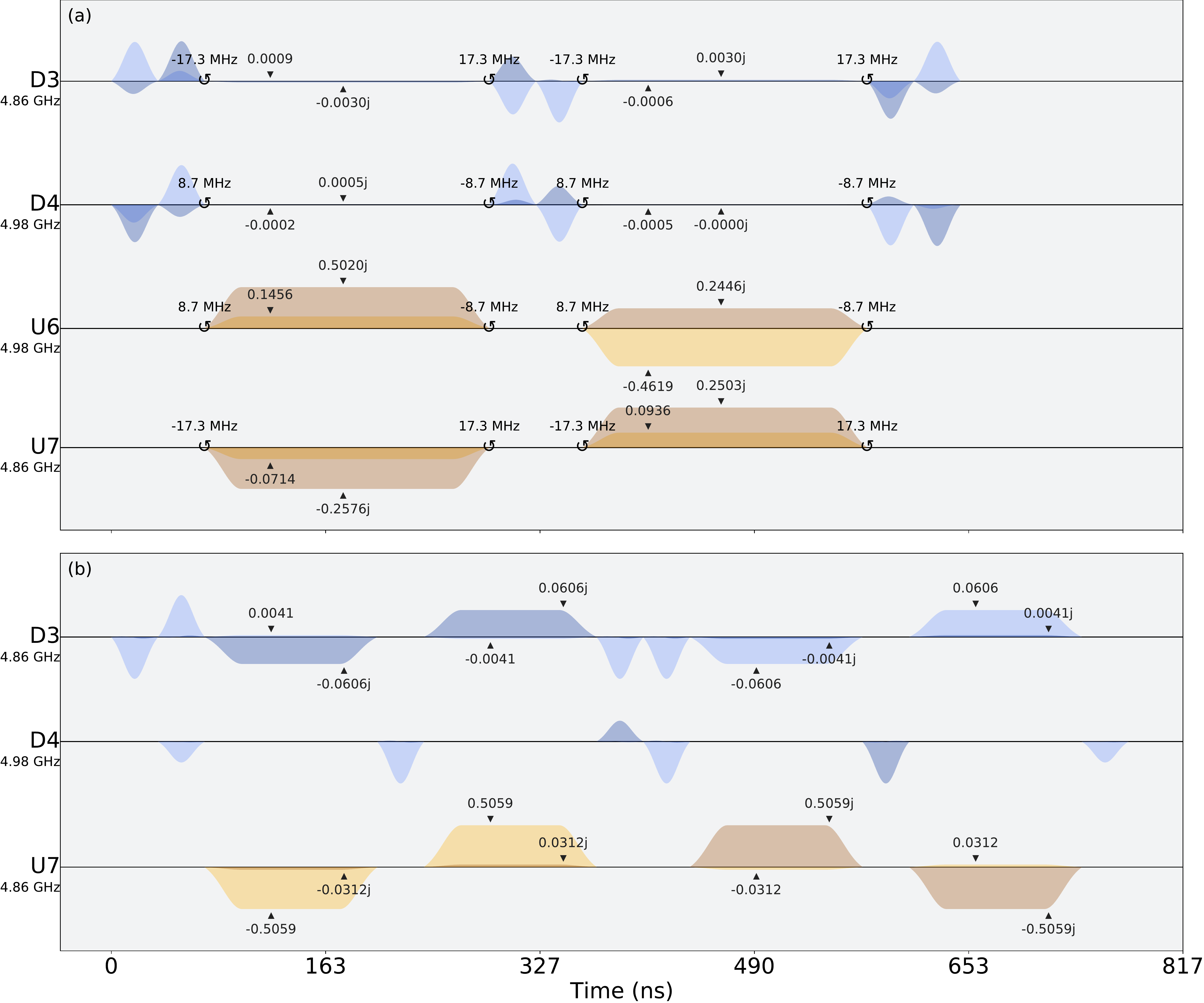}
		\caption{
		Pulse schedules of iSWAP gate with different implementations; (a) CCR gates and (b) TPCX gates.
		Each horizontal line corresponds to a pulse channel associated with a qubit with a specific frame.
		Chanel names are shown on the left side with the frequency assigned to the channel.
		Curly arrows indicate a frequency shift of the channel.
		These frequency shifts are introduced to correct drive induced Stark shift.
		Pulse amplitude of real and imaginary part of CCR and CR gates are also shown with triangle signs.
		}
		\label{waveform_iswap}
	\end{center}
\end{figure*}

\Cref{waveform_iswap} shows the comparison of calibrated pulse \texttt{Schedule}s of the iSWAP gate with different implementations.
The key ingredient of the CCR gate is the bidirectional CR drive on \texttt{ControlChannel} U6 (U7), which applies microwave tones to Q3 (Q4) in the frame of Q4 (Q3).
The weak active cancellation tones are simultaneously applied to the \texttt{DriveChannel}s D3 and D4 which apply microwave tones in the frame of associated qubits.
Local operations are realized by two consecutive $\sqrt{\mathrm{X}}$ pulses and three virtual-$Z$ gates on D3 and D4 prepended and appended to CCR gates.
On the other hand, the standard iSWAP decomposition generates a pulse sequence based on the echoed CRs, yielding a slightly longer schedule.
In this decomposition, CR gates can be implemented with the rotary tone~\cite{PRXQuantum.1.020318}, which appears in D3 and D4 in parallel with the CR tones with comparable amplitude to $\sqrt{\mathrm{X}}$ gates.

\end{document}